\documentclass{resonance}

\usepackage{times}
\usepackage{graphicx}
\usepackage{fancybox}
\usepackage{amsmath}
\usepackage{amssymb}
\usepackage{float}

     {\begin{Sbox}\begin{minipage}}%
     {\end{minipage}\end{Sbox}\fbox{\TheSbox}}

\begin{document}

\title{Gravitational collapse and structure formation in an expanding
  universe with dark energy} 

\author{Manvendra Pratap Rajvanshi, Tuneer Chakraborty \&  J. S. Bagla}

\maketitle

\authorIntro{\includegraphics[width=2cm]{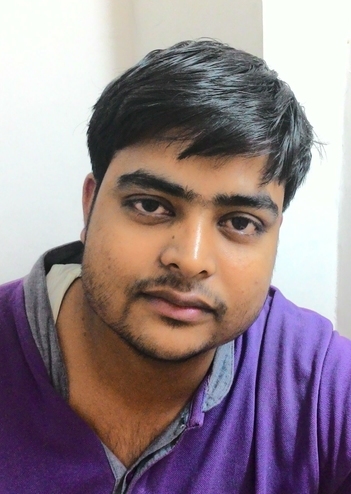}\\
  Manvendra is doing PhD at IISER Mohali with research
  interests in structure formation, dark energy and computational 
  methods.  He is studying 
  gravitational collapse in dark energy models. \\
\includegraphics[width=2cm]{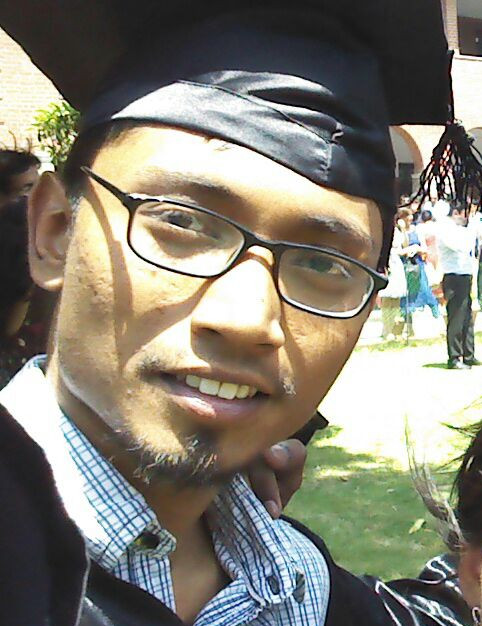}\\
  Tuneer is doing Masters in astrophysics from TIFR
  Mumbai. He currently works in helioseismology  
and is also interested in General Relativity. \\
\includegraphics[width=2cm]{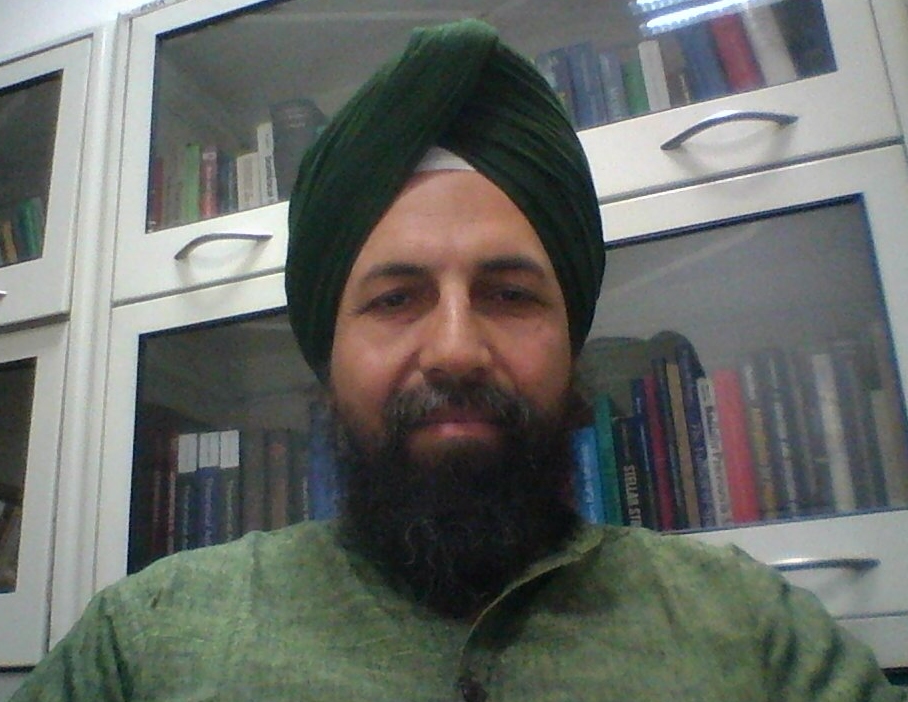}\\
Jasjeet works at IISER Mohali.  He is interested in diverse problems
in physics, his research is in cosmology and galaxy formation.}

\begin{abstract}
  Observations show that the expansion of the Universe is accelerating.
  This requires that the dominant constituent of matter
  in the Universe has some unusual properties like negative pressure.
  This exotic component has been given the name {\sl dark energy}. 
  We work with the simplest model of dark energy, the cosmological
  constant introduced by Einstein.
  We study the evolution of spherical over-densities in such a model
  and show that there is a minimum over-density required for collapse:
  perturbations with a smaller amplitude do not collapse.
  This threshold is interesting as even perturbations with a positive
  over-density and negative energy do not collapse in finite time.
  Further, we show that perturbations with an amplitude larger than,
  but comparable to the threshold value, take a very long time to
  collapse.   
  We compare the solutions with the case when dark energy is absent. 
\end{abstract}

\monthyear{month year}
\artNature{GENERAL  ARTICLE}

\section{Introduction}

Ever since Newton formulated his theory of gravitation, there has
been a quest for a mathematical theory for large-scale dynamics of
the Universe.
At cosmological scales, gravity is the dominant force, hence the
theory of gravitation is central to study of Universe.
The early discussion in this context focused on the question of
stability of the large-scale distribution of matter, and whether the
Universe is required to be infinite in extent for stability
\cite{1976SciAm.235b..90C}.

Present day models of the Universe are based on relativistic
theories of gravity.  
Einstein proposed a static cosmological model soon after proposing
the general theory of relativity (see \cite{1917SPAW.......142E} as
well as \cite{O’Raifeartaigh2017} and references therein).  
He invoked the cosmological term, now known as the cosmological
constant in order to obtain a static model. 
It was soon shown that the static model is unstable to
perturbations, i.e., if there are any fluctuations in density then
the under-dense regions expand and over-dense regions undergo rapid
collapse leading the universe away from the static equilibrium.

Over the next fifteen years, many scientists tried to construct 
relativistic cosmological models.
Friedmann \cite{fr_met,2013ASPC..471...71B}, Lemaitre
\cite{lem_hub,1931MNRAS..91..483L}, Robertson
\cite{rob_hub,1935ApJ....82..284R} and Walker
\cite{1935QJMat...6...81W} arrived at the simplest relativistic
cosmological models. 
These models assumed a universe that is homogeneous and isotropic. 
This assumption was an extension of the Copernican principle that
there is no preferred location or direction in the Universe.
The models can be described as different manifestations of a
space-time metric that is now called the
Friedmann-Lemaitre-Robertson-Walker (FLRW) metric. 
A common feature of solutions in all the models was an expanding
universe, which was shown to match observations made by Slipher
\cite{1915PA.....23...21S}, and, Hubble \cite{1929PNAS...15..168H}
and Humason.
In any region, it is expected that galaxies are seen to move away
from each other due to expansion, and the recession velocities are
larger for galaxies at a higher distance.
\begin{displaymath}
  V = H_0 R
\end{displaymath}
where $R$ is the distance from an observer to a galaxy, $V$ is the
recession speed, and $H_0$ is the Hubble's constant and its observed
value is close to $70$~km.s$^{-1}$.Mpc$^{-1}$.
This relation is valid for distances that are much smaller than
$c/H_0 \simeq 4286$~Mpc.
Here Mega Parsec (Mpc) is a unit of distance and equals $3.08 \times
10^{22}$~m.
The Hubble's constant is dimensional and also gives us a natural
time scale of $\sim 1.4 \times 10^{10}$~years. 
For more details about the Hubble's law and its discovery, see
\cite{2009Reson..20..803B}. 

The inclusion of a cosmological constant in the equations describing 
the Universe leads to interesting consequences.
A cosmological constant can lead to accelerated expansion of the
Universe.  
The cosmological constant has been used at many times to alleviate
the age problem: without the cosmological constant the age of the
Universe is constrained to be smaller than $\tau_0 = H_0^{-1}$,
where $H_0$ is the Hubble's constant.
Earliest measurements of the Hubble's constant were incorrect and
yielded $\tau_0 \simeq 1.8$~billion years (with
$H_0=550$~km.s$^{-1}$.Mpc$^{-1}$), smaller than the age of the oldest
rocks on the Earth\footnote{See
  https://en.wikipedia.org/wiki/Age\_of\_the\_Earth for discussion of the
determination of the age of oldest rocks by Arthur Holmes.}. 
The measurements of the Hubble's constant were corrected over three
decades when several subtle problems with the original measurements
were discovered and corrections were made.
For example, the
differences between population I and population II stars were
realized by Baade \cite{1956PASP...68....5B}.
Population I stars are similar to the Sun in their chemical
composition.
Population II stars have a much smaller amount of elements other than
Hydrogen and Helium.
Astronomers refer to all other elements as metals and the metallicity
of population II stars is much lower than that of population I stars.
Population II stars are found in the halo of the Galaxy.
Baade discovered that Cepheid variables in the two
populations\cite{1956PASP...68....5B} have a different period
luminosity relation and this affects the distance estimates.
Further, it was realized that Hubble and Humason had confused 
H{\sc ii} regions with bright stars and this led to an incorrect
inference of distance. 
It took another four decades for the error bars to shrink below
$10\%$, and by this time it had been shown using observations of
supernovae type Ia\mfnote{Supernovae of type Ia are
  very bright.  These events mark the crossing of the Chandrasekhar
  mass limit for a white dwarf as it gains mass from a companion star.
  The luminosity of supernovae is related to the decline of flux in
  fifteen days.  This permits astronomers to estimate the distance and 
  redshift for these objects independently, and hence constrain the
  rate of expansion.} that the cosmological constant is non-zero and
dominates the energy budget of the
Universe\cite{1998Natur.391...51P,1998AJ....116.1009R,1998ApJ...507...46S}.  
Observations of supernovae of type Ia indicate that the expansion of
the Universe is accelerating, and the cosmological constant is the
simplest model that fits all available observations.
The general term used for the component responsible for an
accelerating universe is {\sl dark energy}, and the cosmological
constant discussed here is the simplest model of dark energy.
As we will see below, dark energy is required to have negative
pressure. 

In this paper we present a discussion of the effect of a
cosmological constant on non-linear evolution of density
perturbations. 
Dynamics in the cosmological model called $\Lambda$CDM, which
incorporates both cold dark matter\mfnote{Cold dark matter (CDM) is a
  class of dark matter where the velocity dispersion of dark matter
  particles is very small, and bulk velocities are also
  non-relativistic.  Observations indicate that most of the dark
  matter is of this type.} and dark energy in form of
cosmological constant has been studied in detail.
It is possible to study non-linear collapse of perturbations
analytically if we restrict ourselves by imposing a symmetry, e.g.,
spherical symmetry on the perturbations.
We assume that the density perturbation is spherically symmetric,
all bulk motions are purely radial, and the matter undergoing
collapse does not have a significant velocity dispersion.
Collapse of perturbations in such a case was first studied by Gunn
and Gott\cite{1972ApJ...176....1G} and this has been reviewed by us
in an earlier article \cite{2015Reson..20..803B}.
We strongly urge the reader to refer to that article first and treat
this article as part II in the series.  
The analysis for spherical collapse has been generalized to the case
where a cosmological constant is
present\cite{1984ApJ...284..439P,1991MNRAS.251..128L,1993MNRAS.262..717B,
  1996MNRAS.282..263E}. 
We follow the approach proposed by Barrow and Saich
\cite{1993MNRAS.262..717B} in this paper. 
Basic equations describing the Universe are introduced in \S{2},
these are modified to describe spherically symmetric perturbations
in \S{3} where we also discuss initial  conditions.
Presence of a cosmological constant leads to a critical threshold
for the initial density contrast: smaller density perturbations do
not collapse.
This is discussed in section \S{3} and \S{4}.
Detailed evolution of perturbations is discussed in \S{4} and \S{5}.  

\section{$\Lambda$ in FLRW Equations} \label{sec1}

Einstein's theory of general relativity provides the framework for
studying dynamics of the Universe.
This is also the appropriate framework for studying evolution of
large-scale perturbations.  
However, in the context of spherically symmetric perturbations and
only non-relativistic matter other than the cosmological constant, the
general relativistic equations have a well defined Newtonian limit.
The usual force equation is modified with the addition of a repulsive
interaction term.
Following \cite{2015Reson..20..803B}, we write the equation of motion
for a thin spherical shell at a distance $R$ from the centre.
The choice of origin is arbitrary in case of a smooth universe. 
\begin{equation}
  \ddot{R} = - \frac{GM}{R^2} + \frac{\Lambda}{3} R
\end{equation}
The first term on the right represents gravitational attraction due to
the mass enclosed in the shell, and the second term represents the
repulsion due to the cosmological constant $\Lambda$.
The cosmological constant does not depend on location in space or on
time. 
We can calculate the first integral from equation~(1) with the assumption
that the mass $M$ within the shell is a constant:
\begin{equation}
  \frac{1}{2} {\dot{R}}^2 - \frac{GM}{R} - \frac{1}{6} \Lambda R^2 =
  \mathrm{constant} = \alpha
\end{equation}
The constant here plays the role analogous to energy in particle
dynamics. 
The shell may expand or contract, all other modes of motion lead to
departures from homogeneity and isotropy\mfnote{Homogeneity is the
  translation symmetry, implying that the distribution of matter is
  the same everywhere at a given time.  Isotropy is the invariance
  under rotation, i.e., the matter distribution is the same in all
  directions.} and hence are not admissible in this model.
Using comoving coordinates $\mathbf{r}$ (see, e.g.,
\cite{2015Reson..20..803B}) $\mathbf{R}(t) = a(t) \mathbf{r}$, where
the physical coordinate 
$\mathbf{R}$ may change with time but $\mathbf{r}$ remains fixed for
fundamental observers in an expanding/contracting homogeneous and
isotropic universe, i.e., there is a group of observers whose only
motion in physical coordinates is expansion or contraction and this
can be described entirely in terms of an overall expansion or
contraction of the Universe with no other component to their motion. 
The expansion and contraction is described by the scale factor
$a(t)$.

By expressing the mass $M$ in terms of the average matter density
$\bar\varrho$, 
and changing over from physical coordinate $\mathbf{R}$ to comoving
coordinate $\mathbf{r}$, equations~(1-2) can be written as:
\begin{eqnarray}
  \frac{\ddot{a}}{a} &=& - \frac{4}{3} \pi G \bar\varrho +
  \frac{1}{3}\Lambda  \\
  \left(\frac{\dot{a}}{a}\right)^2 + \frac{k}{a^2} &=&  \frac{8 \pi
                                                       G}{3}
  \bar\varrho + \frac{1}{3}\Lambda
\end{eqnarray}
Here, $k$ is related to the constant $\alpha$ in equation~(2).
By convention, it takes values $0$, $\pm 1$. 
A comparison with the equations in \cite{2015Reson..20..803B} shows
that the cosmological constant counters deceleration in expansion due
to matter.
Matter density is diluted during expansion\mfnote{Matter is conserved
  and so $\varrho a^3 = $~constant.  In other words, $\varrho \propto
  a^{-3}$.}, however the cosmological 
term is a constant and its effect does not suffer a corresponding
reduction.
Thus we expect the cosmological constant to play an important role at
late times in an expanding universe as it can overtake matter as
the dominant source in Friedmann equations. 

Of special interest here is a spatially flat universe ($k=0$) as this
is consistent with observations \cite{2018arXiv180706209P}.
In this case, the Friedmann equation takes the form:
\begin{equation}
H^2 =    \left(\frac{\dot{a}}{a}\right)^2 =  \frac{8 \pi G}{3} \bar\varrho +
   \frac{1}{3}\Lambda = H_0^2 \left[\Omega_M
     \left(\frac{a_0}{a}\right)^3 + \Omega_\Lambda\right] 
\end{equation}
Here, $a_0$ is the present value of the scale factor, $H_0$ is the
present value of the Hubble parameter $H = \dot{a} / a$, $\Omega_M$
and $\Omega_\Lambda$ are the density parameters for matter and
cosmological constant, respectively\mfnote{Density parameter is the
  density in units of the critical density, see
  \cite{2015Reson..20..803B} or any textbook on cosmology for a
  complete definition.}.   
For a flat universe, we have $\Omega_M + \Omega_\Lambda = 1$. 
The solution to this equation gives the scale factor as a function of
time:
\begin{equation}
  \left(\frac{a}{a_0}\right)^3\,=\,\frac{\Omega_{M}}{\Omega_{\Lambda}} \sinh^2 
  \left(\frac{3tH_0\sqrt{\Omega_{\Lambda}} } {2} \right)     
\end{equation}
When $tH_0 \ll 1$, i.e., at early times, this matches with the
evolution of scale factor for a matter dominated universe with $a(t)
\propto t^{2/3}$. 
At late times, $t H_0 \gg 1$, we get accelerated expansion with $a(t)
\propto \exp(2 tH_0\sqrt{\Omega_{\Lambda}})$.
The rate of expansion begins to accelerate when $\ddot{a} > 0$, i.e.,
\begin{eqnarray}
  - \frac{4}{3} \pi G \bar\varrho +
  \frac{1}{3}\Lambda &=& - \frac{1}{2} H_0^2 \Omega_M
  \left(\frac{a_0}{a}\right)^3 + H_0^2 \Omega_\Lambda > 0 \nonumber \\
  \implies \frac{a}{a_0} &>& \left( \frac{\Omega_M}{2\Omega_\Lambda}
  \right)^{1/3} \nonumber
\end{eqnarray}
In our Universe, the transition to accelerating expansion happened
when the scale factor was slightly less than $60\%$ of its present
value.
Note that this epoch is distinct from the one where we transition from
a matter dominated to a dark energy dominated universe: this
transition happens when the scale factor is close to $75\%$ of its
present value.
Thus accelerated expansion starts well before the energy density in
the Universe is dominated by dark energy. 

\section{Spherical Over-density}\label{sec2}

We can consider a spherical over-density in a flat universe ($k=0$).
We consider a shell with radius $R$. 
Due to the over-density, the shell does not correspond to a fundamental
observer and its comoving coordinates change in time, i.e., it
will depart from the fundamental observer located at the same radius at
the initial time. 
The shell contains mass $M$ and average density $\varrho =
\left(1 + \bar\delta\right)\bar\varrho$, where $\bar\varrho$ is the
average density of the Universe as defined earlier and $\bar\delta$ is
the average density contrast inside the shell. 
We choose initial conditions such that $R_{in} \propto a_{in}$ and
the shell is comoving with the Hubble expansion, e.g., ${\dot{R}}_{in}
= H_{in} R_{in}$.
This, along with the Friedmann Equation for the background universe
allows us to compute the constant in equation~(2). 
\begin{equation}
\alpha = \frac{1}{2}{\dot{R}}_{in}^2 - \frac{GM}{R_{in}}
  - \frac{1}{6} \Lambda R_{in}^2 = - \frac{1}{2} H_{in}^2
  \Omega_M(a_{in}) R_{in}^2 {\bar\delta}_{in} \simeq - \frac{1}{2}
  H_{in}^2  R_{in}^2 {\bar\delta}_{in}
\end{equation}
Here, $\Omega_M(a_{in})$ is the density parameter for matter at the
initial epoch.
The only approximation that has been made here in the last step is
that the initial conditions are set at an epoch where matter dominates
over dark energy, i.e., $\Omega_M(a_{in}) \simeq 1$. 
If the region is over-dense then the constant is negative, however
this does not necessarily mean that the shell will undergo collapse.
To see this, we consider the condition for the expanding shell to
decelerate to rest and begin collapsing, which means that $\dot{R}$
must go to zero.
\begin{equation}
  -\frac{GM}{R} - \frac{1}{6} \Lambda R^2 = - \frac{1}{2} H_{in}^2
  R_{in}^2 {\bar\delta}_{in} 
\end{equation}
This is a cubic equation in $R$ and it has a real and positive
solution only if the value of ${\bar\delta}_{in}$ is above a certain
threshold value $\delta_{thr}$ for a given set of cosmological parameters.
The threshold density contrast is independent of the scale of the
perturbation.
We can express $\delta_{thr}$ as the solution to the following equation: 
\begin{equation}
  27\Omega_\Lambda (1+\delta_{thr})^2 a_{in}^3 - 4\Omega_{M} (\delta_{thr} a_0)^3 = 0
  \label{eq12}
\end{equation}
If we set up initial conditions at $a_0/a_{in}=100$ then $\delta_{thr}
\simeq 0.02 $ for $\Omega_\Lambda=0.7$.
Perturbations with $0 < {\bar\delta}_{in} \leq \delta_{thr}$ are over
dense but will never collapse to form a halo.

The critical threshold for density contrast obtained here is very
different in character to a similar threshold obtained in an open
model ($k=-1$ in equation~(4)) in that the constant
$\alpha$ is negative for all positive ${\bar\delta}_{in}$ for the
models under consideration.
As $\alpha$ is related to the energy of the shell, this seems to
indicate that the shell is bound even though it does not collapse in
finite time.
In an open model for the Universe, the threshold density contrast
corresponds to $\alpha=0$, implying that perturbations with a lower
density contrast are not bound.

It is important to note that even for a closed model, the same
criterion leads to a threshold, except that in this case the threshold
is a finite negative value for ${\bar\delta}_{in}$ implying that even
some under-dense regions may undergo collapse.

Thus there are two regimes for perturbations that do not collapse in
presence of dark energy: 
\begin{itemize}
\item
  Energy of the perturbation is non-negative.  This is true only if
  the density contrast is negative.  The perturbation in this case is
  clearly not bound and will not collapse.
\item
  Energy of the perturbation is negative, but the density contrast is
  less than or equal to the threshold value required for collapse in
  finite time, i.e., $0 \leq {\bar\delta}_{in} \leq \delta_{thr}$.  In this
  case the perturbation is bound (negative energy) but will not
  collapse in finite time.  In this case dark energy dominates the
  evolution of the perturbation.  
\end{itemize}
This is to be contrasted with the open models where the energy is
non-negative only if the density contrast is positive and greater than
zero, and such perturbations do not collapse.

\subsection{Equations and Solutions} \label{sec1p1}
  
Initial conditions allow us to obtain the following equation for
dynamics of a shell
\begin{equation}
  \dot{R}^2\,=\, \Omega_{\Lambda}R^2  +
  \Omega_{M}\left(\frac{a_0}{a_{in}}\right)^3\frac{(1+{\bar\delta}_{in}) R_{in}^3}{R}
  - \Omega_{M} \left(\frac{a_0}{a_{in}}\right)^3 \delta_{in}
  R_{in}^2  \label{eq7} 
\end{equation}
where $\delta_{in}$ is initial density contrast and dot represents
derivative with respect to $tH_0$, with $H_0$ being the Hubble's
constant. 
At any time $t$ the density contrast can be obtained from following relation:
\begin{equation}
  (1+\bar{\delta})\, =
  \,(1+{\bar{\delta}}_{in}) \left(\frac{a}{a_{in}}\right)^3
  \left(\frac{R_{in}}{R}\right)^3 
  \label{eq8}       
\end{equation}
This relation follows from conservation of mass within the shell.
The factor $1 + {\bar{\delta}}_{in}$ accounts for the initial over
density within the shell, $(a_{in}/a)^3$ is the evolution of the
average density of the universe that is the background in which the
over density is evolving, and $(R_{in}/R)^3$ is the change in density
of the shell due to change in its radius.  
  
Solution to equation~(\eqref{eq7}) can be obtained in the form of an
integral (Hyper-geometric function) which has to be evaluated numerically.
Instead one can also solve the differential equation~(\eqref{eq7})
numerically and obtain the solution.
Here we present the solution and contrast it with solution when there
is no $\Lambda$, i.e., with the Einstein-deSitter case. 
   
\begin{figure}[!t]
    \caption{Solutions for radius of over-density as a function of time
      for four different values of 
      $\frac{{\bar\delta}_{in}}{\delta_{thr}}$. Dash-dot lines are for
      $\Lambda CDM$ while continuous lines are for Einstein-deSitter
      case.
      For over densities less than
      the critical threshold value the collapse happens in
      Einstein-deSitter case but not in cosmological constant
      model. Even for perturbations 
      above the threshold, perturbations collapse in presence of
      cosmological constant but these take more time to collapse as 
      compared to Einstein-deSitter.}
    \includegraphics[width=4truein]{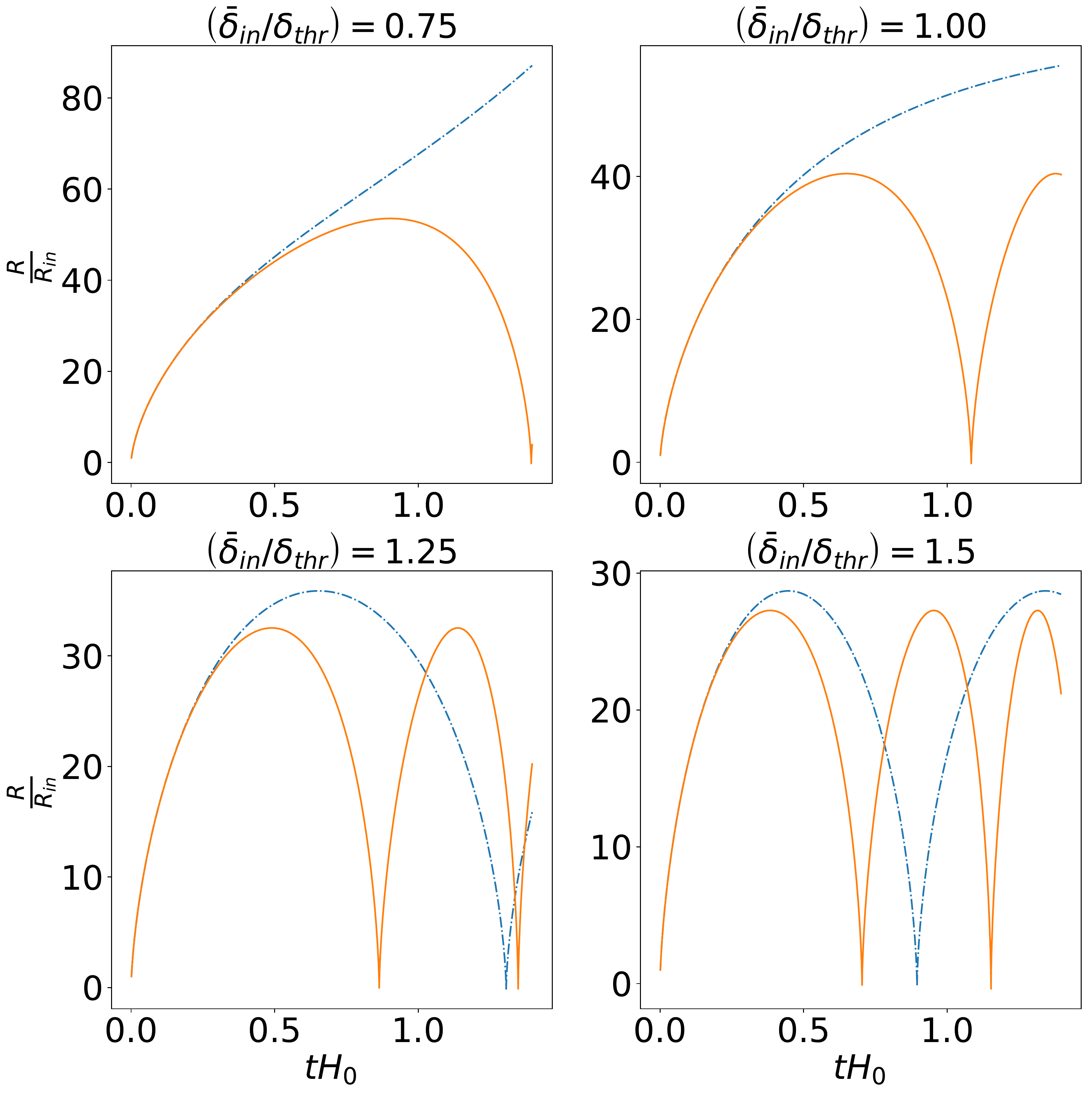} \label{fig1}
\end{figure}

We plot solutions of equation~(\ref{eq7}) in Figure~1.
In this figure we compare the solutions with the same initial
conditions for a universe with cosmological constant with the
corresponding solutions in the Einstein-deSitter universe.
We have plotted four cases around ${\bar\delta}_{in}/\delta_{thr} =
1$, the threshold initial value that divides collapsing and
non-collapsing solutions.
We find that in the Einstein-deSitter universe, all over-densities
lead to collapse whereas in presence of cosmological constant, only
regions with ${\bar\delta}_{in} > \delta_{thr}$ lead to collapse.
Further, in case where cosmological constant is present, collapse
takes much longer time than the corresponding over-density in an
Einstein-deSitter universe.
The difference between the two diminishes as we go to
${\bar\delta_{in}}/\delta_{thr} \gg 1$. 
We also note that the radius at which the shell begins to recollapse
is larger in the presence of dark energy.

\section{Turn Around}\label{sec3}

The initial conditions are set such that the shell is expanding with
the Universe.
For the shell to collapse and form a bound structure, the expansion of
the shell must come to a halt and the radial velocity must vanish at a
finite time. 
This stage of zero radial velocity is called {\sl turn around} where
the perturbation reaches its maximum radius.  
Whether we reach such a stage or not depends on the competition
between attractive gravitational force and repulsion due to
cosmological constant. 
Setting ${dR}/{d(tH_0)}=0$ in equation~(\eqref{eq7})  we obtain
an expression for turn around radius by solving the equation: 
\begin{eqnarray}
  R_{ta}\,&=&\,\frac{3(1+{\bar\delta}_{in})}{{\bar\delta}_{in}} R_{in}
  \left(\frac{4\Omega_{M}({\bar\delta}_{in} a_0)^3} 
  {27\Omega_{\Lambda}(1+{\bar\delta}_{in})^2 a_{in}^3}
  \right)^{1/2}
  \nonumber \\
  &&  \sin\left[ \frac{1}{3}
  \arcsin\left\{\left(\frac{27\Omega_{\Lambda}(1+{\bar\delta}_{in})^2  
  a_{in}^3}{4\Omega_{M}({\bar\delta}_{in} a_0)^3}\right)^{1/2} \right\}
  \right] \label{eq10}
\end{eqnarray}
For the solution to be physically meaningful, the argument of
$\arcsin$ in curly brackets must be less than unity.  
This is the origin of the condition given in equation~(9) that the
initial density contrast must be larger than the threshold value
$\delta_{thr}$ for the perturbation to reach turn around.
 
\begin{figure}[!t]
  \caption{The combination
    $R_{T}\delta_{in}/R_{in}$ is plotted as a function of
    ${\bar\delta}_{in}/\delta_{thr}$.  We choose this combination as its
    expected value is unity in the Einstein-deSitter model.  We see
    that the turn around radius is larger in presence of the
    cosmological constant and its value increases as the initial
    density contrast ${\bar\delta}_{in}$ approaches $\delta_{thr}$ from
    above.}
  \includegraphics[width=2.9truein]{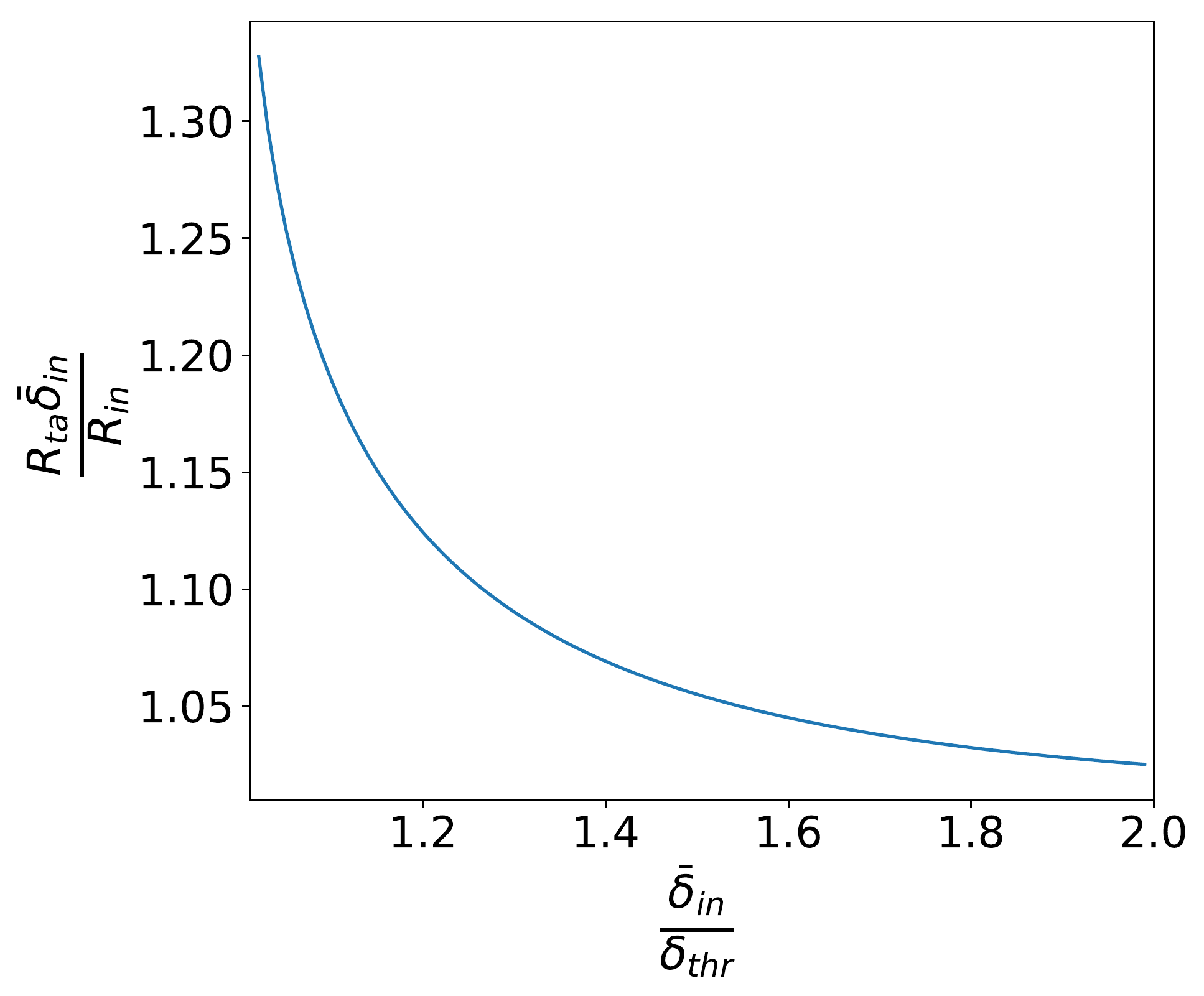} \label{fig2}
\end{figure}
      
\subsection{Characteristics at turn around}

The radius of the perturbation at turn around is fixed uniquely by the
initial over-density.
In case of the Einstein-deSitter universe, the radius of the shell
increases by a factor of $1/{\bar\delta}_{in}$ from the initial
value, thus $R_{ta} = R_{in}/{\bar\delta}_{in}$.
Thus a perturbation with a larger initial over-density expands by a
smaller factor before turning around and collapsing. 
In Figure~2 we show the combination $R_{ta}{\bar\delta}_{in}/R_{in}$,
which is unity in the Einstein-deSitter universe.
We plot this combination as a function of
${\bar\delta}_{in}/\delta_{thr}$.
We find that $R_{ta}{\bar\delta}_{in}/R_{in}$ tends to unity as
${\bar\delta}_{in}/\delta_{thr}$ becomes much larger than unity.
However, as we approach the threshold value ${\bar\delta}_{in}/\delta_{thr}
= 1 $ from above, the turn around radius becomes larger than the
corresponding perturbation in the Einstein-deSitter universe.
We can conclude from here that the density of the perturbation at turn  
around is lower than the corresponding perturbation in the
Einstein-deSitter universe. 

\begin{figure}[!t]
  \caption{It takes more time for less dense (smaller
    ${{\bar\delta}_{in}}/{\delta_{thr}}$) perturbations to reach the
    maximum radius (turn around) and hence by the time it reaches
    turn around, the background density has decreased more and hence
    we see a higher density contrast at turn around for these
    smaller initial densities (see previous figure).}
  \includegraphics[width=2.9truein]{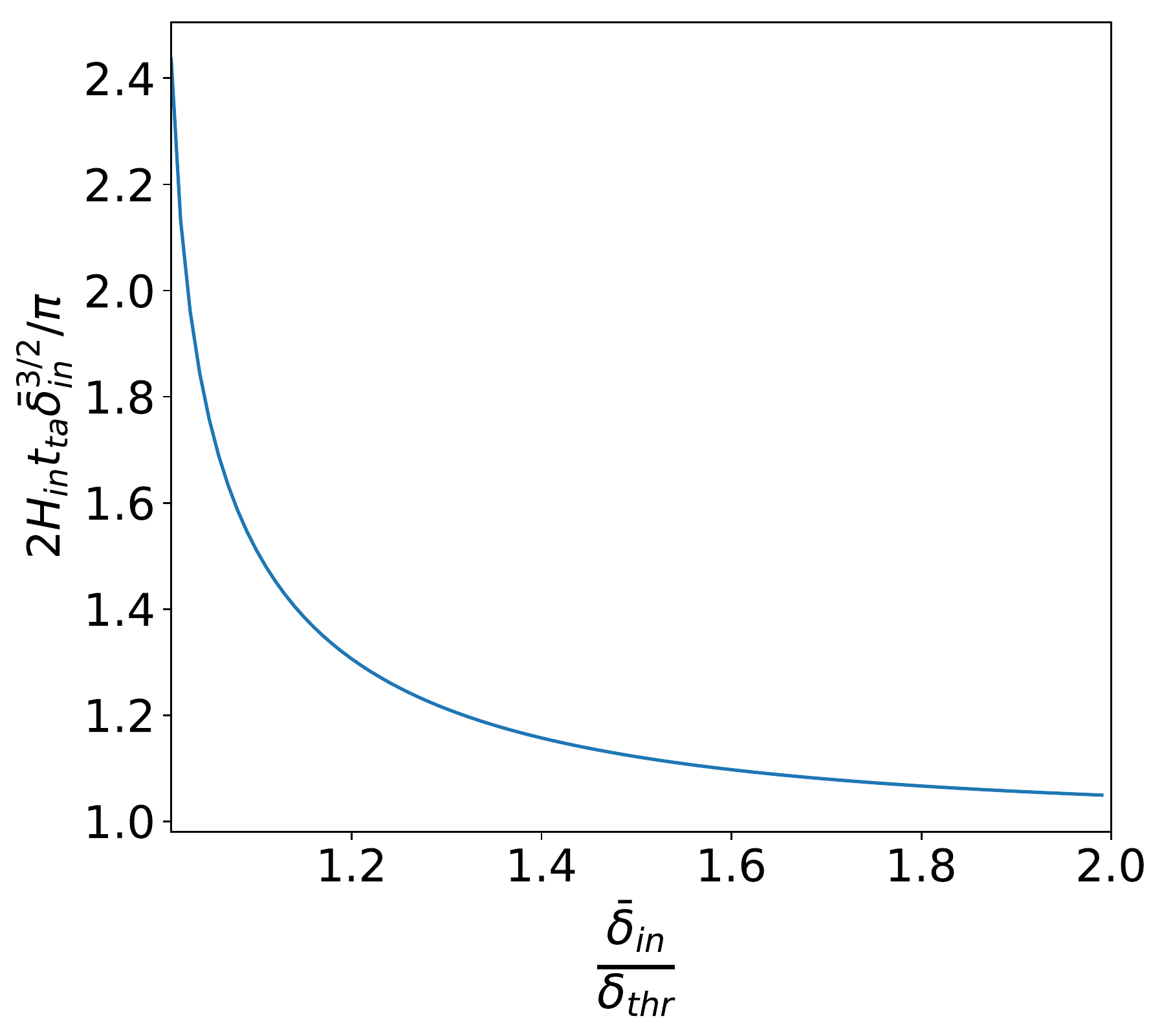} \label{fig3b}
\end{figure}

In Figure~3, we show the time taken to reach turn around.
In Einstein-deSitter universe we have $t_{ta} = \pi/(2 H_{in}
{{\bar\delta}_{in}^{3/2}})$, implying that a perturbation with a
larger initial over-density takes less time to reach turn around.
We have plotted $2 H_{in} t_{ta} {{\bar\delta}_{in}^{3/2}}/
\pi $ as a 
function of ${\bar\delta}_{in}/\delta_{thr}$.
We find that this also approaches unity in the limit
${\bar\delta}_{in}/\delta_{thr} \gg 1$. 
However, we see that the combination becomes large and diverges as we
approach  ${\bar\delta}_{in}/\delta_{thr} = 1 $ from above.  
This indicates that for perturbations near the critical threshold,
time taken to reach turn around is significantly larger than the
corresponding perturbations in the Einstein-deSitter universe.
Lower panels in Figure~1 also illustrate this point. 

Lastly, we plot the density contrast at turn around, 
${\bar\delta}_{ta}$, as a function of ${\bar\delta}_{in}/\delta_{thr}$.
This is shown in Figure~4. 
In the reference model, i.e., the Einstein-deSitter model, this has
the value $9 \pi^2 / 16 -1 \simeq 4.55$.
We find that the density contrast at turn around in the case of a
universe with the cosmological constant is {\sl larger} and the value
diverges as we approach  ${\bar\delta}_{in}/\delta_{thr} = 1 $ from above.
We have already noted above that the density at turn around is
smaller, and we can reconcile these statements by noting that as
perturbations take longer to reach turn around in a universe dominated
by a cosmological constant, the average density of the Universe drops
to lower values by the time the perturbation reaches turn around.
Thus we have a larger density contrast or over-density at turn around,
but smaller density of matter in the perturbation!

\begin{figure}[!t]
  \caption{Density contrast at turn around is plotted as a function of
    ${\bar\delta}_{in}/\delta_{thr}$.  The expected value of density contrast
    at turn around in the Einstein-deSitter model is $4.55$.  We see
    that the density contrast at turn around in presence of
    cosmological constant is higher, and increases sharply as
    ${\bar\delta}_{in}/\delta_{thr}$ approaches unity from above.  The
    perturbations are larger at turn around when the cosmological
    constant is present but the density contrast is higher as it takes
    longer to reach turn around, and the average density of the
    universe decreases by a large factor in this time.}
  \includegraphics[width=2.9truein]{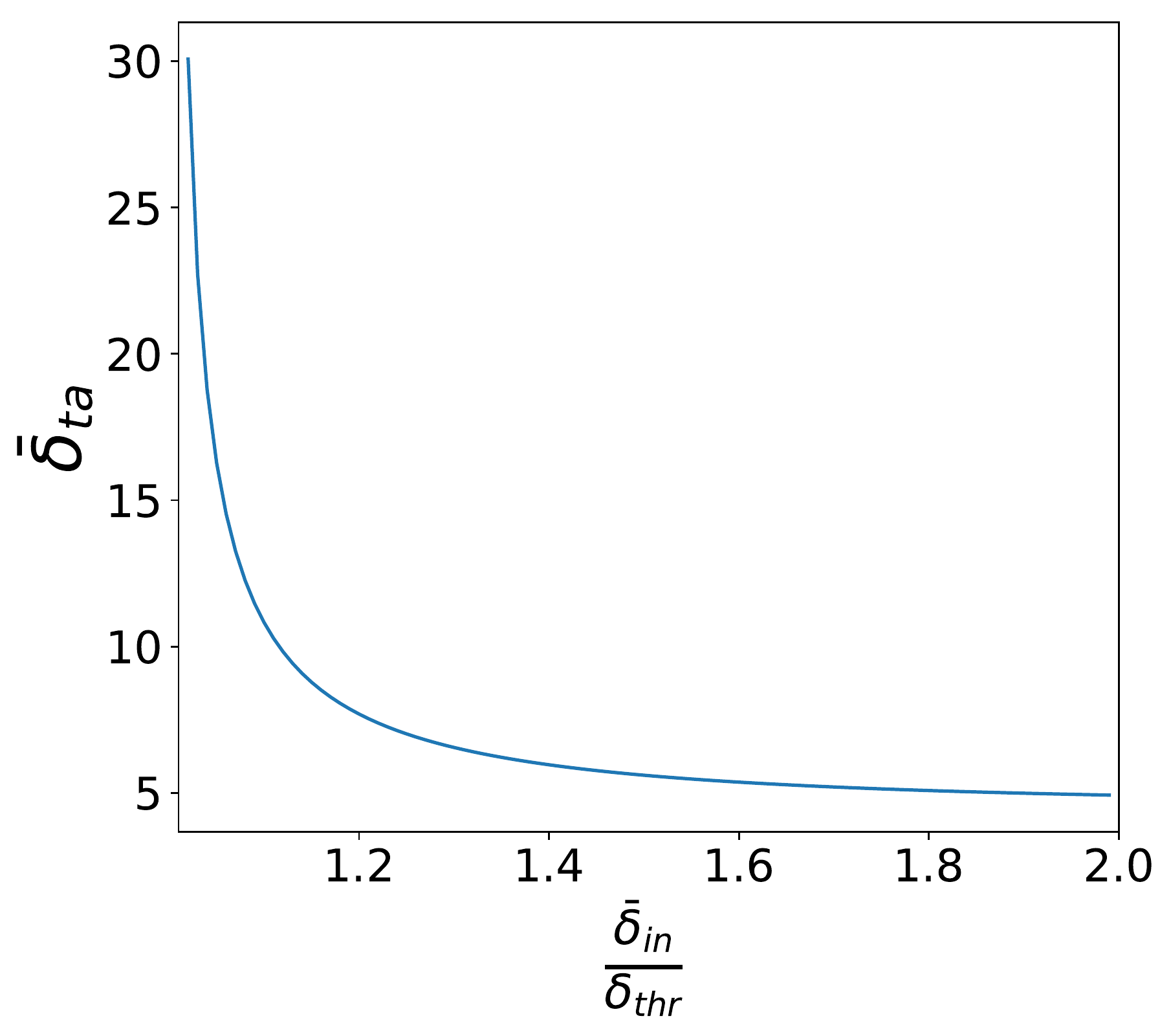} \label{fig3}
\end{figure}

\section{Virialization} \label{sec4}

Solutions for shells in cases where we have eventual collapse are
oscillatory in nature, as illustrated in Figure~1.
With such a solution, we may expect each shell to expand, turn around
and then collapse to the origin before bouncing again.
Such a fate is avoided in a real physical system where small
non-radial motions arising from a small initial velocity dispersion
are amplified during collapse and rapidly lead to an equilibrium state
within the time it would have taken the shell to collapse back to the
origin.
The dynamical equilibrium is achieved when the system virializes.
Radius of the system in equilibrium can be calculated using the
virial theorem\mfnote{Dynamical equilibrium is achieved for a bound
  system when the average motions of particles in the system
  counter-balance the interaction forces.  The motions of particles
  are characterized by the average kinetic energy and the interaction
  forces are characterized in terms of the average potential energy.
  Given the form of interaction, the virial theorem provides a
  relation to be satisfied in dynamical equilibrium.}. 
\begin{eqnarray}
\langle 2 T + \vec{F}.\vec{r}_{vir} \rangle &=& 0 \nonumber\\
 \mathrm {where }\,\,\, \vec{F} &=& -{\vec\nabla}U \nonumber  \\
 \therefore \langle 2 T -\vec{r}.{\vec\nabla} U \rangle &=& 0
 \nonumber 
\end{eqnarray}
where $T$ is kinetic energy and $U$ is potential energy.
In presence of the cosmological constant, there are two contributions
to the potential energy and we need to take both into account.
Further, given that we are discussing a single shell, we may drop the
averaging to avoid confusion.  
For matter, gravitational potential is proportional to $r^{-1}$, and
for $\Lambda$ it is proportional to $r^2$.
Therefore,
\begin{eqnarray}
\left(\vec{r}.\vec{\nabla}\right) U_m &=& -U_m \nonumber \\
\left(\vec{r}.\vec{\nabla}\right) U_{\Lambda} &=& 2U_{\Lambda} \nonumber
\end{eqnarray}
Thus, we find that at virial equilibrium:
\begin{equation}
  2T + U_m - 2U_{\Lambda}= 2 \left(T + U_m + U_\Lambda \right) -
  U_m - 4
U_\Lambda = 2 E - U_m - 4 U_\Lambda = 0
\end{equation}
The expression in brackets is the constant we evaluated in equation~(7).
We may evaluate it at turn around as at that stage the kinetic energy
vanishes. 

We use this relationship to determine virial radius in $\Lambda$CDM model.
Since at turn around the kinetic energy term is zero and because net
energy is conserved, we have at Virial radius: 
\begin{eqnarray}
&&E =  - \frac{GM}{R_{ta}}  - \frac{1}{6} \Lambda R_{ta}^2 \nonumber
\\
\implies &&  - 2 \frac{GM}{R_{ta}} - \frac{1}{3} \Lambda R_{ta}^2 +
\frac{GM}{R_V} + \frac{2}{3} \Lambda R_V^2 = 0 \nonumber \\
\implies && 2 \frac{GM}{R_{ta}} \left( 1 -
\frac{R_{ta}}{2 R_V}\right) = - \frac{1}{3} \Lambda R_{ta}^2 \left(1
- 2 \left(\frac{R_V}{R_{ta}}\right)^2 \right) 
\end{eqnarray}
It is clear that in the case of $\Lambda=0$, we get $R_V =
R_{ta}/2$.
In the general case, substituting for $M$ and $\Lambda$, we get:
\begin{equation}
\begin{split}
R_V = &\left(\frac{2}{3}\right)^{1/2} \left( \frac{\Omega_\Lambda
  R_{ta}^3+\Omega_M (\frac{a_0}{a_{in}})^3 (1+{\bar\delta}_{in})
  R_{in}^3}{\Omega_\Lambda R_{ta}}  \right)^{1/2}  \\ & \sin\left[
  \frac{1}{3} \arcsin \left\{ \frac{\Omega_M 
    a_0^3 (1+{\bar\delta}_{in}) R_{in}^3}{a_{in}^3 R_{ta}^3}
   \left( \frac{1.5}{1  + \frac{\Omega_M}{\Omega_\Lambda} (\frac{a_0
         R_{in}}{a_{in} R_{ta}})^3 (1+{\bar\delta}_{in}) }
   \right)^{3/2}\right\}
   \right]   
\end{split}
\end{equation}  
Figure~5 shows $R_V / R_{ta}$ as a function of ${\bar\delta}_{in}/\delta_{thr}$.
It can be seen that as we go towards larger initial density contrast,
the ratio of virial radius to the turn around radius approaches $0.5$
from below.
We see in Figure~2 that the turn around radius $R_{ta}$ is larger in
presence of dark energy.
Combining these, we infer that the virial radius in models with dark
energy is higher as compared to models without any dark energy. 
The density of collapsed halos is lower as a result,
though the density contrast is higher as collapse takes longer and the
density of the background falls to lower values in this time.

\begin{figure}[!t]
  \caption{The ratio of virial radius to the turn around radius is
    shown as a function of ${\bar\delta}_{in}/\delta_{thr}$.  As
    ${{\bar\delta}_{in}}/{\delta_{thr}}\rightarrow \infty $ the curve tends
    to Einstein-deSitter limit of $0.5$.}
  \includegraphics[width=2.9truein]{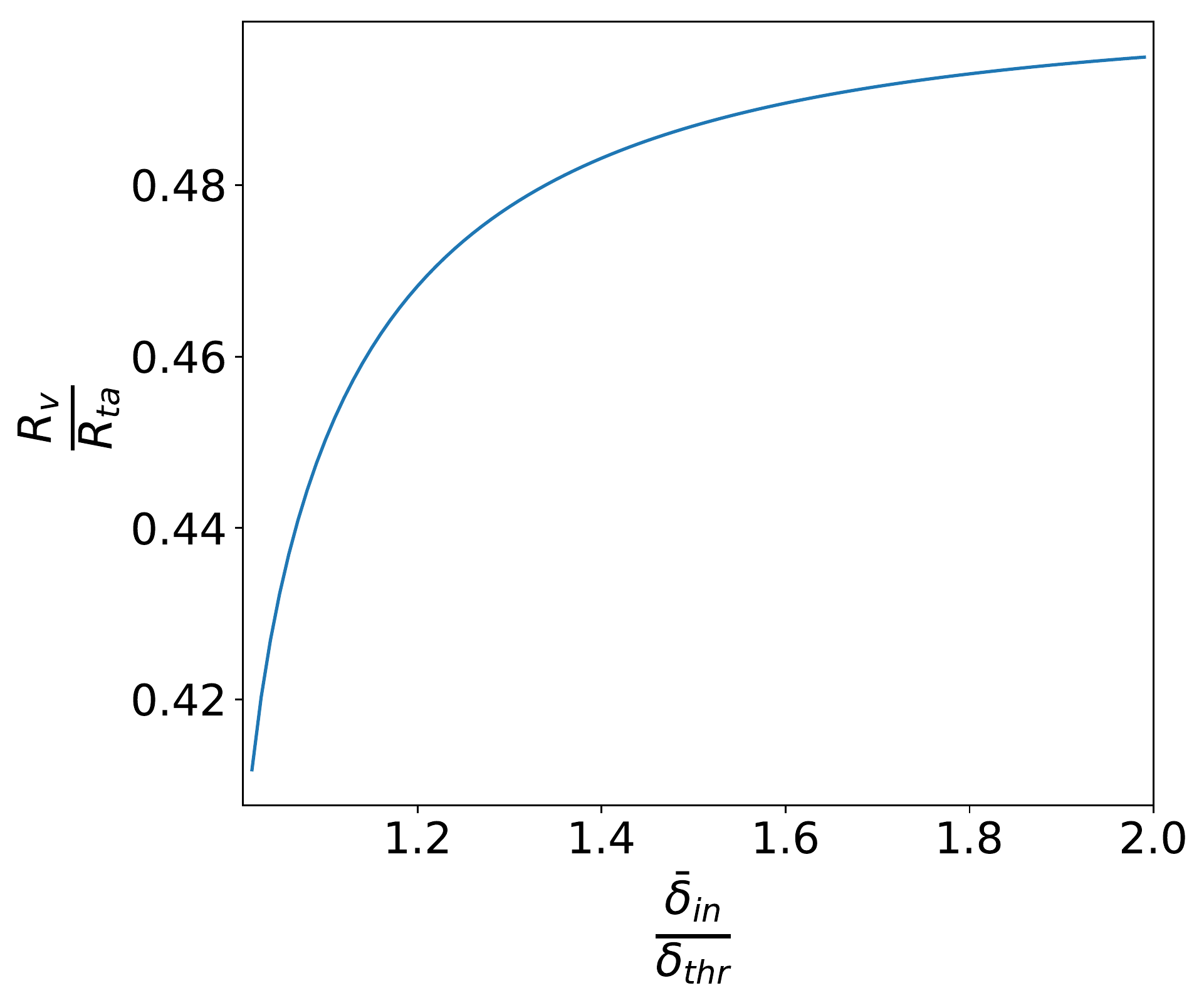} \label{fig4}
\end{figure}

The density contrast at virialization is shown in Figure~6, again as a
function of ${\bar\delta}_{in}/\delta_{thr}$.
We find that the density contrast is very high for perturbations where
${\bar\delta}_{in}/\delta_{thr}$ is just over unity.
This is primarily because these perturbations take a very long time to
reach virialization and the Universe expands by a larger factor in
this time.
This is illustrated in Figure~7.
Here we plot the scale factor at virialization time ($a_V/a_0$) for
different initial perturbations.
Here, $a_0$ is the present value of the scale factor.  
Time of virialization ($t_V$) is generally taken to be $2t_{ta}$ which
is the exact time for analytical solution (of Einstein-deSitter) to
shrink to zero, i.e., the crossing time.
This continues to be approximately true in presence of the
cosmological model.

\begin{figure}[!t]
  \caption{Density contrast at virialization ${\bar\delta}_{V}$ as a
    function of  ${\bar\delta}_{in}/\delta_{thr}$.  Density contrast is very
    high for ${\bar\delta}_{in}/\delta_{thr}$ close to unity and it decreases
    as we get to larger initial density contrast, asymptotically
    approaching the expected value for Einstein-deSitter model as
    ${{\bar\delta}_{in}}/{\delta_{thr}}\rightarrow \infty $.} 
  \includegraphics[width=2.9truein]{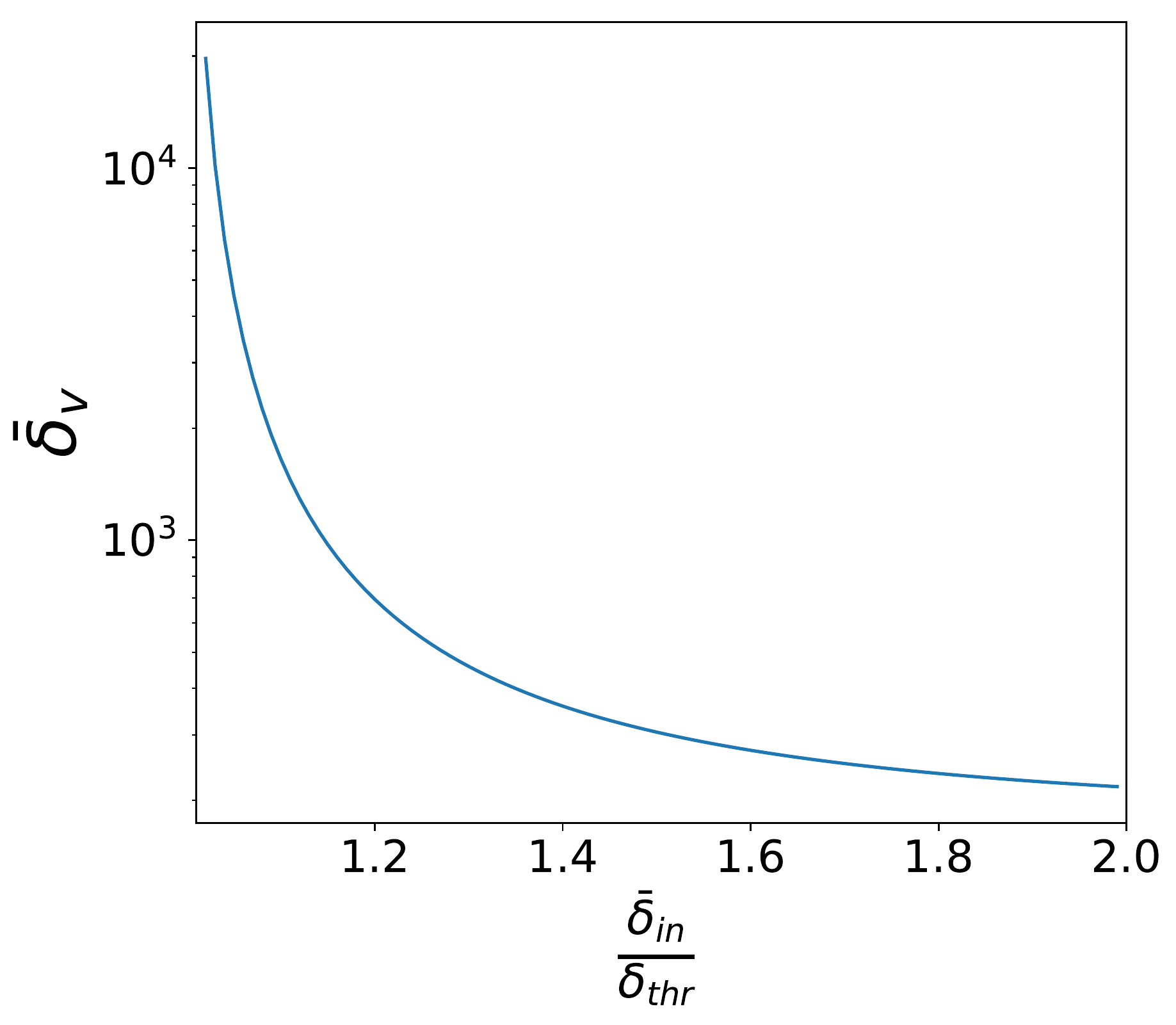} \label{fig5}
\end{figure}

\section{Discussion}

We find that the presence of cosmological constant leads to
introduction of a threshold density contrast: perturbations with a
smaller density contrast do not stop expanding and hence never
collapse.
The regions are over-dense, remain over-dense with a steadily
increasing density contrast that never collapse.
The threshold density contrast is independent of the scale of
perturbation. 

\begin{figure}[!t]
  \caption{We have plotted the scale factor at the time of
    virialization $a_V$ as a function of
    ${\bar\delta}_{in}/\delta_{thr}$.
    It takes more time for perturbations with a smaller
    initial density contrast (${{\bar\delta}_{in}}/{\delta_{thr}}$) to
    virialize and 
    hence by the time it virializes, the background density has
    decreased more and hence we see a higher virial density contrast
    for these smaller densities(see previous figure).} 
  \includegraphics[width=2.9truein]{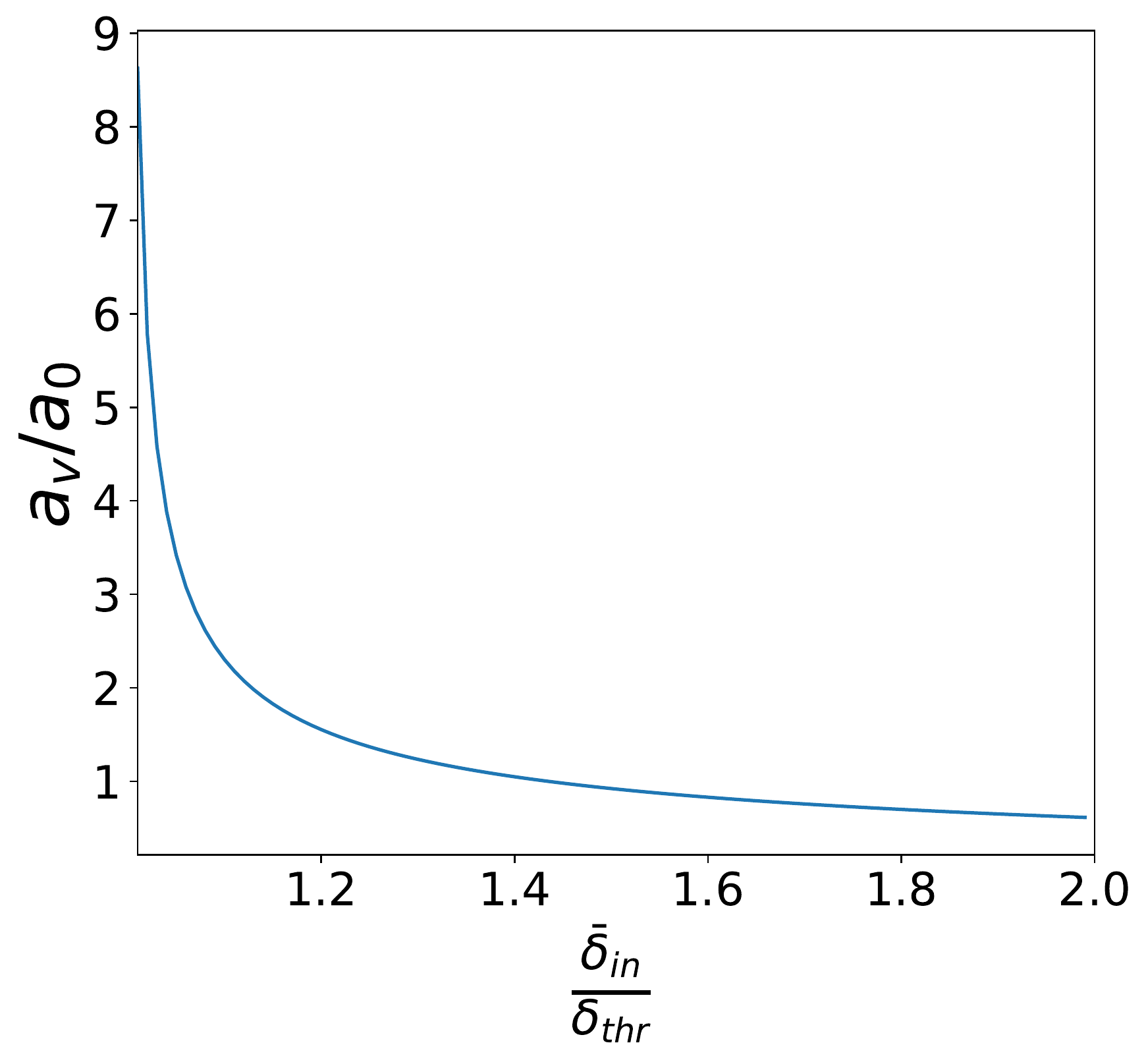} \label{fig6}
\end{figure}

It takes longer to form collapsed structures, hence the process
of structure formation must start with higher initial density contrast
as compared to the Einstein-deSitter universe.
Thus over-densities must have a higher value at early times. 
For example, we expect to see more clusters of galaxies at high
redshifts in such a universe.

These findings have observable implications for galaxy formation and
the formation of large scales structure in the Universe. 

\section*{Acknowledgements}

The authors thank the anonymous referee for a very detailed reading
and helpful comments.

\end{document}